# Spin photogalvanic effect in two-dimensional collinear antiferromagnets

*Rui-Chun Xiao,[1,2,‡] Ding-Fu Shao,[3,‡] Yu-Hang Li,[4,‡] Hua Jiang[1,2]\**

[1]School of Physical Science and Technology, Soochow University, Suzhou 215006, China

[2]Institute for Advanced Study, Soochow University, Suzhou 215006, China

[3]Department of Physics and Astronomy & Nebraska Center for Materials and Nanoscience, University of Nebraska, Lincoln, Nebraska 68588-0299, USA

[4]Department of Electrical and Computer Engineering, University of California, Riverside, California 92521, USA

**ABSTRACT** Spin photogalvanic effect (SPGE) is an efficient method to generate a spin current by photoexcitation in a contactless and ultra-fast way. In two-dimensional (2D) collinear antiferromagnetic (AFM) materials that preserve the combined time-reversal ($\hat{T}$) and inversion ($\hat{I}$) symmetry (*i.e.*, $\hat{T}\hat{I}$ symmetry), we find that the photogalvanic currents in two magnetic sublattices carry different kinds of spins and propagate in opposite direction if the spin-orbit coupling is negligible, resulting in a pure spin current without net charge current. Based on the first-principles calculations, we show that two experimentally synthesized 2D collinear AFM materials, monolayer $MnPS_3$ and bilayer $CrCl_3$, host the required symmetry and support sizable SPGE. The predicted SPGE in 2D collinear AFM materials makes them promising platforms for nano spintronics devices.

Spintronics,[1] which exploits the spin degrees of freedom of electrons in condensed matters for information storage and processing, has generated persistent interests in both physics and engineering. Conventional spintronics devices are usually constructed by three-dimensional magnetic materials, and their application in nanoscale spintronics are hence strongly limited by the sizable thickness. The recent discovered two-dimensional (2D) van der Waals magnets naturally circumvent this limitation, thus bring an opportunity for next-generation nanoscale spintronic devices. In particular, 2D collinear antiferromagnetic (AFM) materials are especially interesting (such as bilayer CrX$_3$ (X=Cl, Br, I),[2-4] even-layer MnBi$_2$Te$_4$,[5, 6] monolayer MnPS$_3$[7, 8]). Except for the advantage of low dimensionality, 2D collinear AFM materials also exhibit ultrafast dynamic response, are robust against external magnetic perturbations, and have no stray field.[9-11] However, despite remarkable progress in exploring exotic phenomena in 2D collinear AFM materials,[2-6, 12] the applications of them so far have been restricted to introduce a ferromagnetic order or a net magnetization. The direct spintronic devices based on the intrinsic AFM order have yet to be determined.

One promising mechanism to directly use these 2D collinear AFM materials is the spin photogalvanic effect (SPGE),[13-21] which generates the spin current by photoexcitation in a contactless and ultra-fast way. In AFM materials, two sublattices with opposite spin polarizations are connected by certain symmetry. If the spin-orbit coupling (SOC) effect is negligible, the photogalvanic currents in two magnetic sublattices carry different kinds of spins and propagate in opposite directions, inducing a pure spin current in total.[21] The spin current here has a long spin diffusion length because spin is a conserved quantity. Moreover, the scattering induced by Joule heating and Oersted fields can also be well avoided.

In this letter, we predict that the 2D collinear AFM materials can be utilized to fabricate the nano spintronic devices via the SPGE. We study the SPGE in 2D collinear AFM materials, and demonstrate that the $\hat{T}\hat{I}$ symmetry (the combination of time-reversal $\hat{T}$ and inversion $\hat{I}$ symmetries) guarantees the photogalvanic current with opposite directions and spin polarizations in two spin sublattices, and results in a pure spin current. As specific examples, we consider the Néel-type monolayer MnPS$_3$ and the A-type bilayer CrCl$_3$, and demonstrate, based on first-principles density-functional theory calculations (see Supporting Information), that these materials support a sizable SPGE.

**Model and candidate materials.** A collinear AFM material is composed of two FM sublattices with opposite magnetizations. These two sublattices are usually connected by a symmetry operation, leading to a zero-net magnetization. One of the most common symmetry operations to compensate for the magnetization is the $\hat{T}\hat{I}$ symmetry, as shown in Figure 1a. In such an AFM material, even though both the time-reversal ($\hat{T}$) symmetry and the space inversion ($\hat{I}$) symmetry are broken, the combined $\hat{T}\hat{I}$ symmetry is still preserved and connects the magnetic sublattices with opposite magnetizations. This symmetry enforces the energy bands of two sublattices are degenerate. In the absence of the SOC effect, the spin is a good quantum number, *i.e.* a conserved quantity. It allows us to investigate the SPGE of the $\hat{T}\hat{I}$ preserved AFM materials in two sublattices with spin-up and spin-down magnetizations separately.[21]

We therefore first describe the photogalvanic effect (PGE) in one spin sublattice. When an electron is pumped from the valence band to the conduction band by a linear polarized light, it generates a shift of wave packet in real space. This "shift vector" can be described by[22-24]

$$R_{nm}^a(\boldsymbol{k}) = \frac{\partial \phi_{nm}(\boldsymbol{k})}{\partial k^a} + r_{nn}^a(\boldsymbol{k}) - r_{mm}^a(\boldsymbol{k}), \qquad (1)$$

where $a$ is Cartesian index, $\phi_{nm}$ is the phase factor of the interband Berry connection and $r_{mm}^a$ is the intraband Berry connection matrix along $a$ direction. The PGE element $\sigma_{abc}$ is determined by[21, 22]

$$\sigma_{abc}(\omega) = -\frac{i\pi|e|^3}{2\hbar^2} \int [d\boldsymbol{k}] \sum_{n,m} f_{nm}(\boldsymbol{k}) R_{nm}^a(\boldsymbol{k}) r_{nm}^b(\boldsymbol{k}) r_{mn}^c(\boldsymbol{k}) \delta[\omega_{mn}(\boldsymbol{k}) - \omega], \qquad (2)$$

where $f_{nm}(\boldsymbol{k}) = f_n(\boldsymbol{k}) - f_m(\boldsymbol{k})$ is the difference of occupation factors, $\hbar\omega_{mn}(\boldsymbol{k}) = \epsilon_m(\boldsymbol{k}) - \epsilon_n(\boldsymbol{k})$ is the difference of band energies, and $\omega$ is the frequency of light. Finally, the induced charge PGE current density **J** is

$$J_a = \sigma_{abc}(\omega) E_b(\omega) E_c(-\omega), \qquad (3)$$

where $a, b, c \in \{x, y, z\}$, and $E_a$ is the electric field of light along the $a$ direction. The shift vector is odd under $\hat{I}$ symmetry, *i.e.* $R_{nm}^a(\boldsymbol{k}) = -R_{nm}^a(-\boldsymbol{k})$. Therefore, the integration in Eq. (2) is enforced to be zero in a centrosymmetric material, leading to the vanishing of PGE. For the spin sublattices shown in Figure 1b and Figure 1c, the broken $\hat{I}$ symmetry allows the PGE in each spin sublattice individually. Since the magnetic moments in each sublattice are in FM order, the PGE charge current is spin-polarized with the tensor $\sigma^\uparrow$ ($\sigma^\downarrow$) for the sublattice with spin-up (spin-down) magnetization. The $\hat{T}$ symmetry converts a spin-up electron to a spin-down one, and the $\hat{I}$ symmetry guarantees the shift vector to be opposite, *i.e.*, $R_{nm}^{a,\uparrow}(\boldsymbol{k}) = -R_{nm}^{a,\downarrow}(\boldsymbol{k})$ (see Figure 1b and Figure 1c). On the other hand, the optical transition possibilities ($\propto r_{nm}^b(\boldsymbol{k}) r_{mn}^c(\boldsymbol{k})$) of two spin states in the same **k** points are equal. As a consequence, in a $\hat{T}\hat{I}$ preserved AFM material, the spin polarizations and directions of the PGE current in different sublattices are enforced to be opposite, *i.e.* $\sigma^\uparrow = -\sigma^\downarrow$. This results in a vanishing charge current, but a finite pure spin current, determined by $\sigma^S = \sigma^\uparrow - \sigma^\downarrow$, which leads to the SPGE in an AFM material with weak SOC effect. It is necessary to emphasize that the SPGE current here has longer spin diffusion length compared to the systems with strong SOC effect because of spin conservation. Therefore, we can construct a spintronic device based on SPGE as shown in Figure 1d: a laser light beam irradiate to a 2D collinear AFM material with $\hat{T}\hat{I}$ symmetry, and generates a pure spin current **J**$^s$ without a charge current **J**$^c$, which can inject the spin current to the attached terminal in a contactless and ultra-fast way.

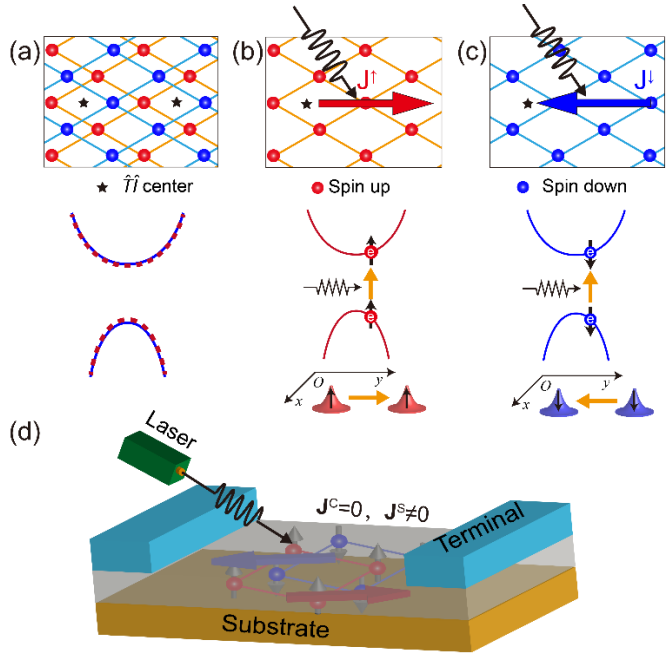

Figure 1. (a) Schematics of an AFM material with $\hat{T}\hat{I}$ symmetry (top) and the corresponding symmetry enforced band structure (bottom). (b,c) Spin-polarized sublattices and the related singlet band structures, respectively. A light generates a PGE current in each sublattice (top), and the corresponding optical transition occurs in the spin subband (bottom). The situation of the sublattice with spin-up and spin-down magnetizations are shown in (b) and (c), respectively. (d) A spintronic device based on SPGE with a 2D collinear AFM material.

There is a rich collection of layered magnetic materials.[25-27] However, not all these magnetic orders support the required symmetry. To determine the suitable AFM order for the SPGE, we use a honeycomb lattice model, a common magnetic structure that most 2D collinear AFM materials adopt. Figure 2 shows four common magnetic orders. The corresponding materials are summarized in the Supporting Information. The zigzag-type and stripy-type AFM orders such as $MPS_3$ (M=Fe, Co, Ni)[28-32] and $MSiSe_3$(M=V, Fe, Ni)[33] are shown in Figure 2a and Figure 2b, respectively. The $\hat{I}$ symmetry of each sublattice can be clearly seen for these AFM orders, preventing the PGE in each sublattice and the total SPGE. Figure 2c shows the Néel-type AFM order, where the adjacent magnetic moments are coupled each other antiparallelly. This AFM order has been found in monolayer $MnPS_3$[7, 8, 34-36] and $MnPSe_3$[37] in experiments. In this AFM order, the $\hat{I}$ symmetry of each sublattice is broken, and two sublattices are connected by the $\hat{T}\hat{I}$ symmetry. This type of AFM order satisfies the symmetry for the SPGE. Different from the above AFM orders existing in the monolayer 2D materials, there is an A-type AFM order that can only emerge in multilayer 2D magnets (Figure 2d), such as $CrX_3$ (X=Cl, Br, I)[2-4] and $MnBi_2Te_4$.[5, 6] In those materials, the magnetic atoms are arranged in FM order within intralayer, and in AFM order between adjacent layers. Although $\hat{I}$ symmetry is preserved for monolayer, it is enforced to be broken in bilayer due to the interlayer stacking, which makes the $\hat{T}\hat{I}$ preserved A-type AFM materials compatible for the realization of the SPGE. Based on the above analyses, we can choose suitable 2D magnetic materials. As specific examples, we take the Néel-type monolayer $MnPS_3$ and the A-type bilayer $CrCl_3$ as two representative materials. The two experimentally synthesized 2D collinear AFM materials both have negligible SOC effects and suitable bandgaps for the SPGE.

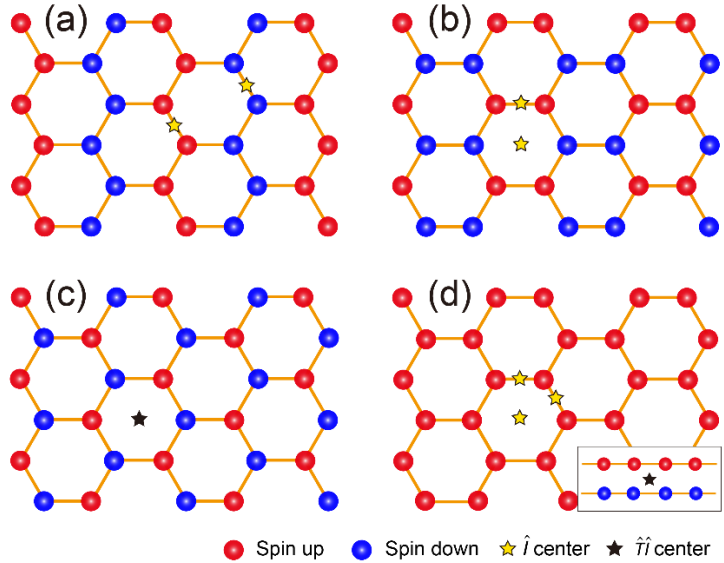

Figure 2. Schematics of four AFM orders in 2D honeycomb lattice. (a) Zigzag-type, (b) stripy-type, (c) Néel-type, and (d) A-type. The inset in (d) shows the side view of A-type AFM order.

**Results and Discussion.** Transition-metal trichalcogenide $MnPS_3$ is an AFM material with a Néel temperature ($T_N$) of 78 K in the bulk.[38, 39] It has a layered structure, where the Mn atoms form a honeycomb lattice in each layer. A dumbbell-shaped $P_2S_6$ unit is located at the center of each Mn hexagon (Figure 3a). The magnetic moments of the nearest-neighbor Mn atoms are antiparallel to each other, forming a Néel-type AFM order. It has been successfully exfoliated down to single-layer recently,[7, 8, 34-36] and the Néel-type AFM order can be maintained. The magnetic sublattices in monolayer MnPS$_3$ have the noncentrosymmetric structure with $D_3$ point group, which contains a $C_3$ rotation symmetry along the $z$-axis, and a $C_2$ rotation axis around the $y$ direction, as shown in Figure 3a. The $\hat{T}\hat{I}$ symmetry that connects two sublattices are clearly reflected in the calculated spin density (Figure 3b). It turns out that the calculated spin density of the non-magnetic P and S atoms are identical for two spin states. Nevertheless, the spin density of the Mn atoms is distributed asymmetrically for the opposite spin states, which leads to the broken $\hat{I}$ symmetry in two spin states. The density of the opposite spin sublattices can be transformed into each other by the application of the $\hat{T}\hat{I}$ symmetry. Figure 3c shows the calculated band structure of monolayer MnPS$_3$, which is consistent with the previous works.[33, 40, 41] A direct bandgap of 2.50 eV is found at the K points.[40] The bands contributed by spin-up and spin-down electrons are degenerate, and also show negligible SOC effect (see Supporting Information). These characteristics support the SPGE in monolayer MnPS$_3$.

There is only one independent tensor element $\sigma_{yyy}^{\uparrow(\downarrow)} = -\sigma_{yxx}^{\uparrow(\downarrow)} = -\sigma_{xxy}^{\uparrow(\downarrow)}$ of the PGE coefficients in each sublattice due to the $D_3$ symmetry (see Supporting Information). The calculated PGE coefficients are consistent with the above symmetry (Figure 3d). The PGE coefficients are zero at low photon energy, and become nonzero when the photon energy is larger than the bandgap. The value of the PGE coefficient can be as high as ~10 $\mu A/V^2$ at 3.36 eV, which is comparable to the reported charge PGE in single-layer monochalcogenides[42-45] and bilayer WTe$_2$.[46] In order to figure out the major contribution of this large PGE coefficient, we project the **k**-solved PGE coefficients of $\sigma_{yyy}^{\uparrow}$ at 3.36 eV to Figure 3e. It can be found that the hot spots (regions where the absolute values of the **k**-solved PGE coefficients are large) appear at the **k** points in ΓM direction and close to the M point. This may originate from the photo transition from the second valence band to the first conduction band near the M point, as indicated in Figure 3b where the valence band shows flat character. In addition, the $\hat{T}\hat{I}$ symmetry ensures the PGE coefficients for the spin-down state are opposite to those for the spin-up state. As a result, the total SPGE coefficients $\sigma^S = \sigma^{\uparrow} - \sigma^{\downarrow}$ are expected to be sizable and can be easily detected under a visible light energy through the magneto-optic Kerr effect or the inverse spin Hall effect. In addition, the SPGE is not limited to the monolayer, because the multilayer MnPS$_3$ also hosts the required symmetry (see Supporting Information).

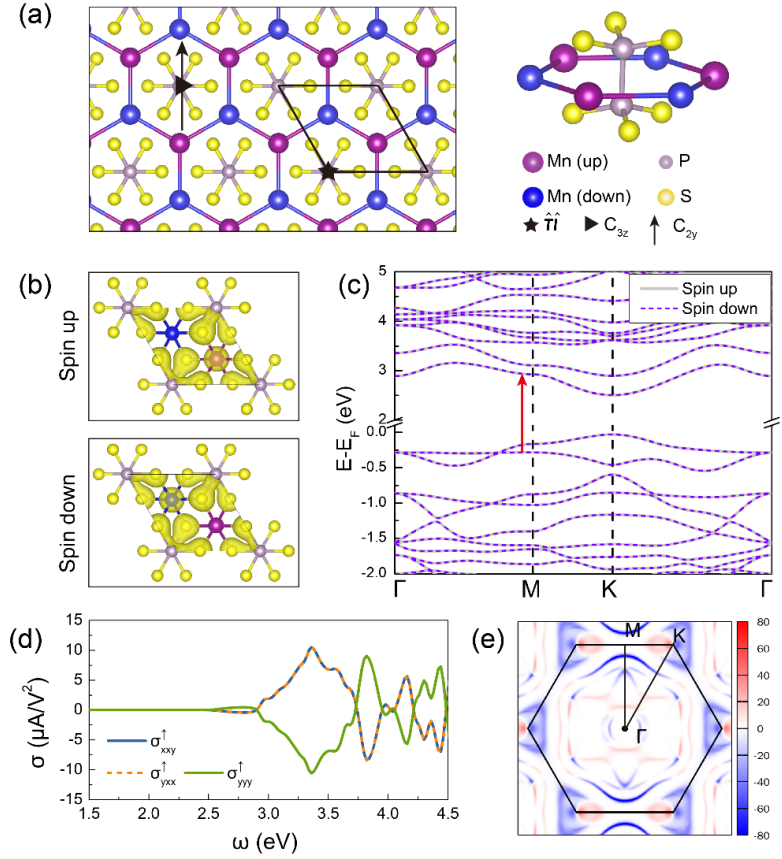

Figure 3. (a) Top view (left) and side view (right) of monolayer MnPS$_3$. The rhombus in denotes the unit cell. (b) Calculated density with the isosurface of 0.047 e/Å$^3$ for spin-up (top) and spin-down (bottom) states. (c) Calculated band structure of monolayer MnPS$_3$. (d) PGE coefficients for the spin-up state. (e) **k**-solved PGE coefficients of $\sigma_{yyy}^{\uparrow}$ (in a.u.) at $\omega =3.36$ eV.

To proceed, we consider the SPGE in bilayer CrCl$_3$. Bulk CrCl$_3$ is an AFM semiconductor with $T_N$=15.5 K.[47] It has been successfully exploited to the atomic limit recently, and the A-type AFM order is maintained down to bilayer.[48-50] In each monolayer, the Cr atoms form a honeycomb structure, and each Cr atom is surrounded by six Cl atoms in octahedral coordination. An interlayer shifting appears when stacking the monolayers (Figure 4a). The magnetic moments of the Cr atoms are ferromagnetically coupled in the same layer while antiferromagnetically coupled between the adjacent layers, forming an A-type AFM order. Although the monolayer CrCl$_3$ is centrosymmetric, the bilayer stacking breaks the $\hat{I}$ symmetry. This broken $\hat{I}$ symmetry is clearly shown in the calculated density for two different spin states, as shown in Figure 4b.

Figure 4c shows the calculated band structure of bilayer CrCl$_3$. A bandgap of 2.61 eV is obtained. The Kramers degeneracy of the band structure results from the $\hat{T}\hat{I}$ symmetry. The sizable PGE coefficients of each spin state are calculated and shown in Figure 4d. There are two independent tensor elements ($\sigma_{xxx}^{\uparrow(\downarrow)} = -\sigma_{xyy}^{\uparrow(\downarrow)} = -\sigma_{yxy}^{\uparrow(\downarrow)}$ and $-\sigma_{yxx}^{\uparrow(\downarrow)} = \sigma_{yyy}^{\uparrow(\downarrow)} = -\sigma_{xxy}^{\uparrow(\downarrow)}$) of the PGE coefficients due to the $C_3$ symmetry. We find that the PGE coefficients are overall smaller than those in monolayer MnPS$_3$, and it is because that the broken $\hat{I}$ symmetry originates from the adjacent layer. However, there is a large peak at ~4 eV for $\sigma_{yyy}^{\uparrow(\downarrow)}$ ($\sigma_{yxx}^{\uparrow(\downarrow)}, \sigma_{xxy}^{\uparrow(\downarrow)}$). The PGE coefficients of spin-down states are opposite to those of spin-up ones due to the $\hat{T}\hat{I}$ symmetry, leading to the SPGE in bilayer CrCl$_3$. A similar SPGE is also expected in even-layer CrCl$_3$. Whereas, for CrCl$_3$ with odd layers, the $\hat{I}$ symmetry is restored, which prevents the SPGE. This indicates an interesting odd-even layer character of the SPGE in few-layer CrCl$_3$.

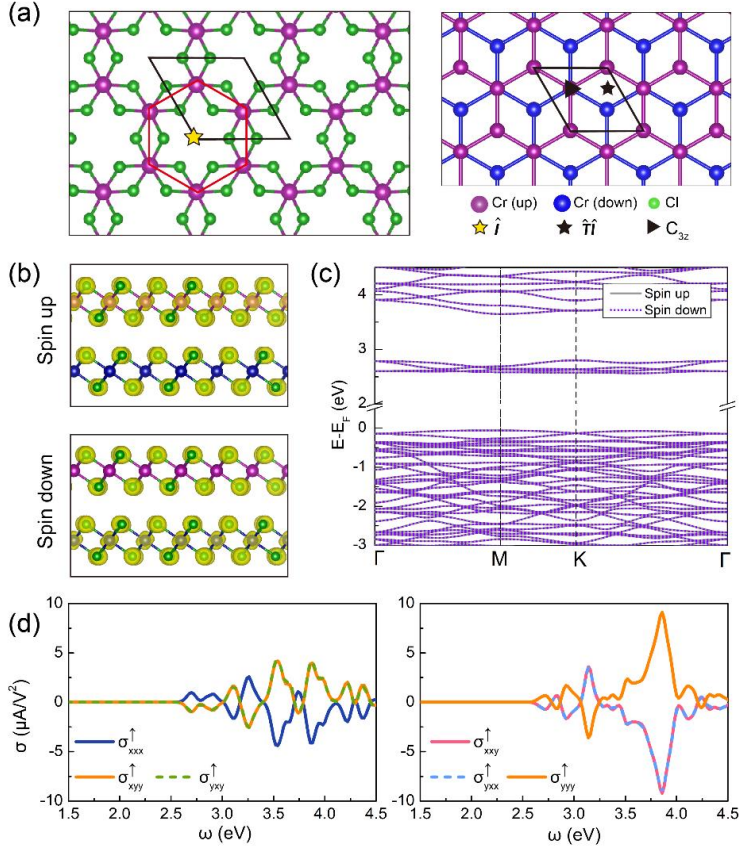

Figure 4. (a) Crystal structure of bilayer CrCl$_3$. The left panel shows the structure of a single layer, and the right panel shows the bilayer stacking (only the Cr atoms are shown). (b) Calculated density (side

views, the isosurface of 0.09 e/Å$^3$) for spin-up (top) and spin-down (bottom) states of bilayer CrCl$_3$. (c) Calculated band structure. (d) PGE coefficients of the spin-up sublattice of bilayer CrCl$_3$.

According to the above calculations, the SPGE enables monolayer MnPS$_3$ and bilayer CrCl$_3$ to be promising candidates for nano spintronics devices, which can inject the pure spin current in a contactless and ultra-fast way. Furthermore, there are many other 2D collinear AFM materials which share similar lattice structures and satisfy the required symmetry of SPGE, such as Néel-type MnPSe$_3$, A-type NiX$_2$, FeX$_2$ (X=Cl, Br), as well as CrBr$_3$, etc. Nevertheless, the materials with strong SOC effects such as A-type CrI$_3$ and MnBi$_2$Te$_4$ are not good candidates for the realization of the SPGE because the spin conservation is heavily violated. However, in these materials, the photoexcitation can still induce charge PGE, which has already been studied recently.[51-53] The SPGE proposed here takes advantage of the intrinsic character of AFM state and also holds great potentials to realize long-distance spin transport. Furthermore, the polarizability of spin current is independent on the photon energy and polarization direction. Finally, we note that the switching Néel vector is equivalent to exchange the positions of the two sublattices, leading to the reverse of SPGE coefficients (see Supporting Information). Therefore, our proposal also offers an efficient method to detect the Néel vector.

**Conclusions.** In summary, we theoretically demonstrate the SPGE in 2D collinear AFM materials with weak SOC, which can be used to generate a pure spin current. Symmetry analyse shows that such SPGE is attributed to two copies of the PGEs with antiparallel polarized spin currents that propagate oppositely in two AFM sublattices. The first-principle calculations demonstrate that two experimental synthesized 2D collinear AFM materials: Néel-type monolayer MnPS$_3$ and A-type bilayer CrCl$_3$ satisfy the symmetry requirement and host detectable SPGE, which are in agreement with the symmetry analysis. These 2D collinear AFM materials can be utilized to fabricate nano spintronic devices in a contactless and ultra-fast way based on SPGE. We hope our work can stimulate further experimental explorations and broaden the research scope of the AFM spintronics.

AUTHOR INFORMATION

**Corresponding Author**

*(H. J.) E-mail: jianghuaphy@suda.edu.cn.

**Author Contributions**

J.H. planned and integrated the research. R.C.X., D.F.S., and H.Y.L. performed the symmetry analysis, and checked the model. R.C.X. preformed the first-principles calculations. The manuscript was written through contributions of all authors. All authors have given approval to the final version of the manuscript. ‡These authors contributed equally.

**ACKNOWLEDGMENT**

We thank Dr. Shu-Hui Zhang and Yang Gao for helpful discussions. This work is supported by the NBRPC under No. 2019YFA0308403, National Nature Science Foundation of China under Grants No. 11947212, No. 11822407, No. 11534001, and Postdoctoral Science Foundation No. 2018M640513.

# Supporting Information

## I. Magnetic structures of AFM materials

Most layer materials have stacking honeycomb structures. They can be divided into two broad families: transition metal halides (including both dihalides and trihalides) and transition metal chalcogenides. Their magnetic structures are summarized in Table S1, and their properties are summarized in the recent review papers.[1-3] According to the symmetry analysis in the main text, monolayer FM, zigzag-type, and stripe-type AFM materials cannot have the spin photogalvanic effect (SPGE) due to processing the inversion symmetry. In contrast, the Néel-type AFM materials and even-layer A-type AFM materials break the inversion symmetry, which are in red in Table S1. Except for the symmetry requirements, the materials should have weak SOC effects. According to the above requirements, we can choose suitable materials.

Table S1 Summary of the structure of 2D magnetic materials.

| Type | FM | A-type AFM | Néel-type | Zigzag-type | Stripy-type |
|---|---|---|---|---|---|
| $MX_2$ | | $FeX_2$ (X=Cl, Br), $CoBr_2$, $NiCl_2$, $NiBr_2$ | | | |
| $MX_3$ | $CrBr_3$, $CrI_3$ | $CrCl_3$, $CrBr_3$, $CrI_3$[4] | | | |
| $MYX_3$ | $CrSiTe_3$, $CrGeTe_3$ | $MnBi_2Te_4$[5, 6] | $MnPS_3$, $MnPSe_3$[7] | $FePS_3$,[8] $CoPS_3$,[9] $NiPS_3$[10] [11] | $VSiS_3$,[7] $VSiSe_3$, $NiSiS_3$, $FeSiS_3$[12] |

## II. Calculation method

The first-principles calculations based on density functional theory (DFT) are performed by using the VASP package. General gradient approximation (GGA) according to the Perdew-Burke-Ernzerhof (PBE) functional are used. The energy cutoff of the plane wave basis is set to 400 eV. The Brillouin zones are sampled with a 12×12×1 mesh of $k$-points. In order to simulate the monolayers, 20 Å vacuum layers are introduced. The vdW force with DFT-D2 correction is considered. Hubbards $U$ term of 3 eV and 5 eV are added for Cr and Mn atoms to account for strong electronic correlations.

The Wannier function method has advantages over the DFT method to calculate the photogalvanic coefficient due to the high-efficiency in the dense k-mesh calculations and the unnecessary of a large number of unoccupied bands.[13] Therefore, the DFT Bloch wave functions are iteratively transformed into maximally localized Wannier functions by the Wannier90 code.[14,15] Mn-$d$ and Se/P-$p$ (Cr-$d$ and Cl-$p$) orbitals are used to construct the Wannier functions for MnPS$_3$ (CrCl$_3$). The effective tight-binding Hamiltonian is obtained to construct the band structures. The PGE coefficients are calculated for spin-up and spin-down bands separately, and numerically evaluated by response theory.[13,16,17] Convergence test of k-mesh is preformed, and 500×500×1 k mesh is sufficient to calculate7 the photogalvanic effect (PGE) coefficients.

The 3D-like PGE coefficients are obtained assuming an active single-layer with the thickness of $L_{active}$[13]

$$\sigma_{3D} = \frac{L_{slab}}{L_{active}} \sigma_{slab},$$

where $\sigma_{slab}$ is the calculated PGE coefficient, and $L_{slab}$ ($L_{active} < L_{slab}$) is the thickness slab.

## III. Symmetry of photogalvanic tensor

Similar to the nonlinear optics tensor, the symmetry of PGE tensor $\sigma^S$ obeys[18]

$$\sigma^S_{ijk} = a_{il} a_{jm} a_{kn} \sigma^S_{lmn},$$

where $S = \uparrow$ or $\downarrow$, and $[a_{il}]_{3\times 3}$ is the symmetry operator matrix. Besides, $\sigma^S$ has the switch symmetry of the last two subscripts, i.e. $\sigma^S_{abc} = \sigma^S_{acb}$. Therefore, the 3×3×3 $\sigma^S_{abc}$ tensor can be contracted as a 3×6 $\sigma^S_{al}$ using the following simplified indices:

$$\begin{array}{cccccccc} bc, cb & \rightarrow & xx & yy & zz & yz, zy & xz, zx & xy, yx \\ l & \rightarrow & 1 & 2 & 3 & 4 & 5 & 6 \end{array}$$

The sub-spin lattice of monolayer MnPS$_3$ has $D_3$ symmetry, so its PGE tensor has the form as

$$\sigma^S_{D_3} = \begin{bmatrix} 0 & 0 & 0 & \sigma^S_{14} & 0 & -\sigma^S_{22} \\ -\sigma^S_{22} & \sigma^S_{22} & 0 & 0 & -\sigma^S_{14} & 0 \\ 0 & 0 & 0 & 0 & 0 & 0 \end{bmatrix}.$$

For 2D materials, $a, b, c \in \{x, y\}$ make sense. Therefore, the elements in red have no physical meaning in this case. Therefore, there is only one independent tensor element and $\sigma^S_{xxy} = \sigma^S_{yxx} =$ $-\sigma_{yyy}^S$ for monolayer MnPS$_3$, which is consistent with the calculation results in the main text. For bilayer and bulk MnPS$_3$, the in-plane $C_3$ rotation symmetry is broken, due to the interlayer stacking. Each sub-spin lattice can be described by $C_2$ (2//y) symmetry, and the corresponding PGE tensor obeys

$$\sigma_{C_2}^S = \begin{bmatrix} 0 & 0 & 0 & \sigma_{14}^S & 0 & \sigma_{16}^S \\ \sigma_{21}^S & \sigma_{22}^S & \sigma_{23}^S & 0 & \sigma_{25}^S & 0 \\ 0 & 0 & 0 & \sigma_{24}^S & 0 & \sigma_{26}^S \end{bmatrix},$$

where the $\sigma_{xxy}^S$, $\sigma_{yyy}^S$, and $\sigma_{yxx}^S$ have no mathematical relationship anymore.

Each sub-spin lattice of bilayer A-type CrCl$_3$ can be described by $C_3$ symmetry, and its PGE tensor obeys

$$\sigma_{C_3}^S = \begin{bmatrix} \sigma_{11}^S & -\sigma_{11}^S & 0 & \sigma_{14}^S & \sigma_{15}^S & -\sigma_{22}^S \\ -\sigma_{22}^S & \sigma_{22}^S & 0 & \sigma_{15}^S & -\sigma_{14}^S & -\sigma_{11}^S \\ \sigma_{31}^S & \sigma_{31}^S & \sigma_{33}^S & 0 & 0 & 0 \end{bmatrix}.$$

There are two independent tensor elements, and $\sigma_{xxx}^S = -\sigma_{xyy}^S = -\sigma_{yxy}^S$, and $\sigma_{yyy}^S = -\sigma_{yxx}^S = -\sigma_{xxy}^S$ for bilayer CrCl$_3$.

## IV. Calculation Results

### A. MnPS$_3$

The calculated lattice constant and bandgap of MnPS$_3$ are shown in Table S2, which show considerable consistent with the experiments. The band structures with SOC effects are shown in Fig. S1, which show weak SOC effects, indicating the SPGE model is suitable for MnPS$_3$.

Table S2 Calculated crystal structure and magnetic moment of MnPS$_3$.

|  | a (Å) | b (Å) | C (Å) | Gap (eV) | Mag ($\mu_B$) |
|---|---|---|---|---|---|
| Monolayer (Calc.) | 6.15 |  |  | 2.50 | 4.61 |
| Bulk (Calc.) | 6.066 | 10.511 | 6.848 | 2.35 | 4.598 |
| Bulk (Expt.) | 6.077 | 10.524 | 6.796 | 2.7[19] | 4.6 [20] |

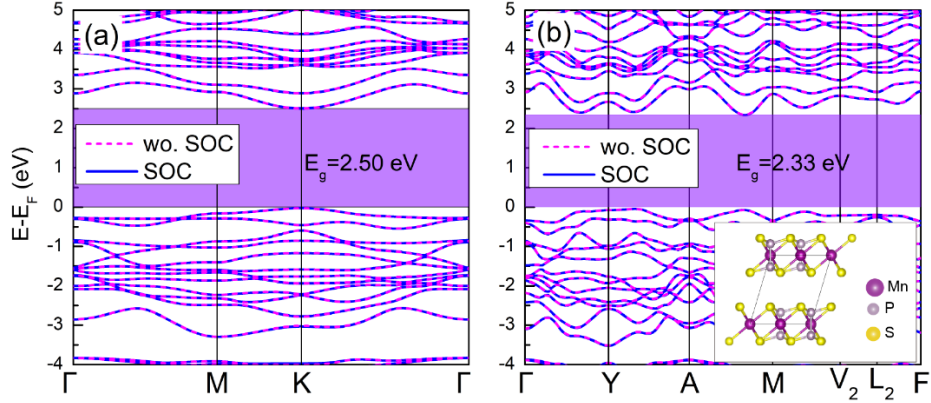

Fig. S1 Band structures of (a) monolayer and (b) bulk MnPS$_3$. The insert in (b) shows the interlayer stacking of bulk MnPS$_3$.

The calculated PGE coefficients of the bilayer and bulk MnPS$_3$ is shown in Fig. S2, which is consistent with the above symmetry analysis. $\sigma^S_{xxy}$, $\sigma^S_{yyy}$, and $\sigma^S_{yxx}$ have no mathematical relationship anymore because the $C_3$ symmetry is broken in the bilayer and bulk MnPS$_3$, yet show weak symmetry relation.

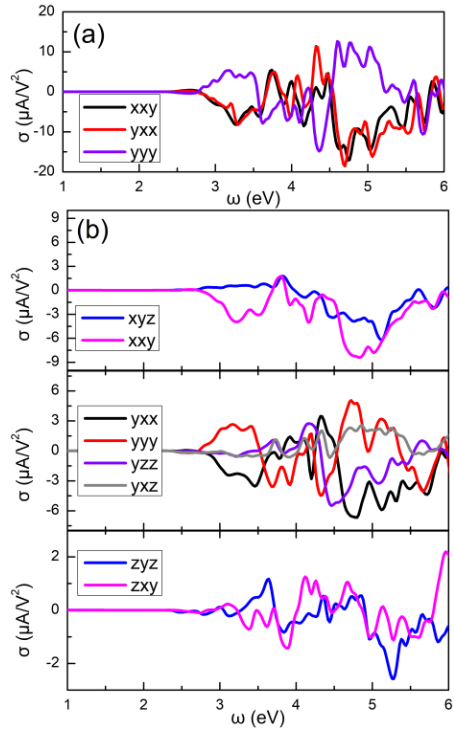

Fig. S2 PGE tensor of (a) bilayer and (b) bulk MnPS$_3$ for spin-up states. The spin-down PGE coefficients are opposite with spin-up ones and are not shown here.

## B. Bilayer CrCl₃

The calculated lattice constants of the bilayer CrCl$_3$ are shown in Table S3. The calculated band structures are shown in Fig. S3, which show weak SOC effects.

Table S3 Calculated crystal structure and magnetic moment of bilayer CrCl$_3$.

|  | a (Å) | Mag ($\mu_B$) | Band gap (eV) |
|---|---|---|---|
| Bilayer (Calc.) | 6.0316 | 3.025 | 2.61 |
| Bulk (Expt. [21]) | 5.9420 |  |  |

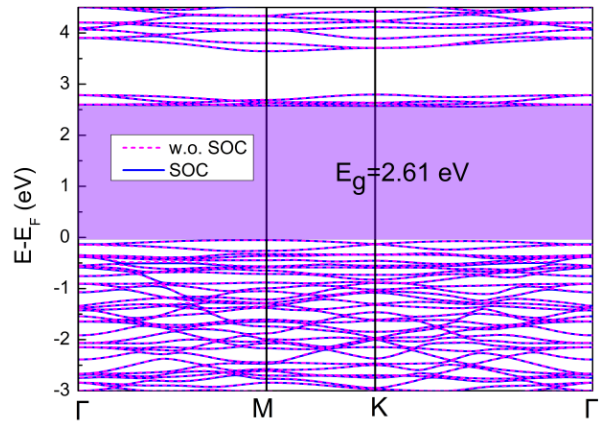

Fig. S3 Band structures of bilayer CrCl$_3$.

## C. Switching the Magnetic order

The switching of magnetic order in MnPS$_3$ and CrCl$_3$ leads to the reverse of SPGE coefficients as shown in Fig. S4 and Fig. S5, respectively.

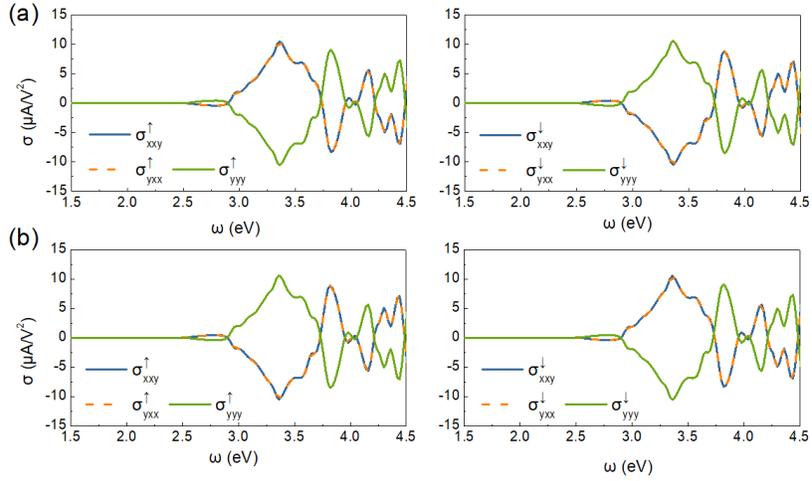

Fig. S4 SPGE coefficients of monolayer MnPS$_3$ with adopting the magnetic order (a) as the main text (b) opposite with the main text.

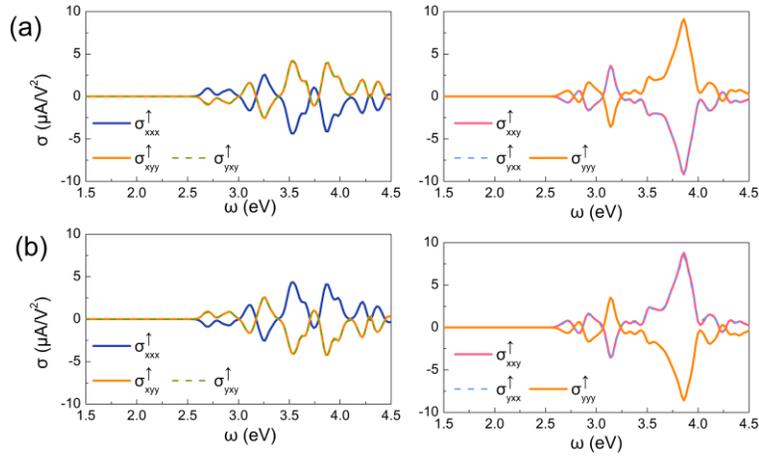

Fig. S5 SPGE coefficients of bilayer CrCl$_3$ with adopting the magnetic order (a) as the main text (b) opposite with the main text. The spin-down PGE coefficients are opposite to spin-up ones and are not shown here.